\documentclass[aps,pra,twocolumn,showpacs]{revtex4}
\usepackage{epsfig}
\usepackage{graphicx}
\usepackage{dcolumn}
\begin{document}
\title{Magic-zero wavelengths of alkali-metal atoms and their applications}
\author{Bindiya Arora}
\affiliation{Department of Physics, IISER Mohali, India}
\author{M. S. Safronova}
\affiliation{Department of Physics and Astronomy, University of Delaware, Newark, Delaware 19716-2593}
\author{Charles W. Clark}
\affiliation{
 Joint Quantum Institute, National Institute of Standards and Technology
and the University of Maryland, Gaithersburg, Maryland 20899-8410, USA}

\date{\today}

\begin{abstract}
Using first-principles calculations, we identify ``magic-zero'' optical wavelengths, $\lambda_{\rm{zero}}$, for which the ground-state
frequency-dependent polarizabilities of alkali-metal atoms vanish. Our approach uses high-precision, relativistic all-order
 methods in which all single, double, and partial triple excitations
of the Dirac-Fock wave functions are included to all orders of perturbation theory. We discuss the use of magic-zero wavelengths for sympathetic
cooling in two-species mixtures of alkalis with group-II and other elements of interest. Special cases in which these wavelengths
coincide with strong resonance transitions in a target system  are identified.

\end{abstract}
 \maketitle

\section{Introduction}~\label{sec1}
The realization of mixtures of trapped ultracold atomic gases ~\cite{2001Sci...291.2570T,PhysRevLett.87.080403,2001Sci...294.1320M,new1,new2} has
opened new paths towards the formation of ultracold diatomic
molecules~\cite{PhysRevLett.97.180404,PhysRevLett.97.120402,2008JPhB...41t3001O,Ni,PhysRevLett.102.020405}, quantum-state control of chemical
reactions~\cite{qstate2010}, prospects for quantum computing with polar molecules~\cite{mille,micheli,2009NJPh...11e5049C}, tests of fundamental
symmetries~\cite{hudson,ray,PhysRevA.80.042508} and studies of fundamental aspects of correlated many-body systems~\cite{tassy}, and dilute quantum
degenerate systems~\cite{hadzi,Roati,regal2,chin,kinast}. Co-trapped diamagnetic-paramagnetic mixtures have also made possible experimental
realization of interspecies Feshbach resonances~\cite{Stan,Inouye,springerlink:10.1140/epjd/e2010-10591-2}, two-species Bose-Einstein condensates and
mixed Bose-Fermi and Fermi-Fermi degenerate gases~\cite{schreck:080403,hadzibabic:160401,Modugno,tassy}.

In an optical lattice, atoms can be trapped in the intensity
maxima or minima of the light field by the optical dipole force~\cite{grimm}.
This force arises from the dispersive interaction of
the induced atomic dipole moment with the intensity gradient of the light field, and is proportional
 to the ac polarizability of the atom.   When its ac polarizability vanishes, as can happen
 at certain wavelengths, an atom experiences no dipole force and thus is unaffected by the presence
 of an optical lattice. Our present work provides accurate predictions of the
``magic-zero wavelengths'' $\lambda_{\rm{zero}}$ which lead to zero Stark shifts for alkali-metal
 atoms. These wavelengths have also been designated as ``tune-out
 wavelengths'' by LeBlanc and Thywissen \cite{LeblancThywissen}.

We suggest some possible uses for such wavelengths, all of which take advantage of the fact
(demonstrated below),
that magic-zero wavelengths are highly dependent upon atomic species and state.
 For a given atomic species and state A,
 let ${\rm L}_{\rm{A}}$ designate an optical lattice or trap made with light at one of the magic-zero wavelengths of A.
We start with a model configuration consisting of the gas A embedded
in ${\rm L}_{\rm{A}}$ and confined by another trap, T.  Some process
is performed on the gas, after which T is turned off.
    Members of A will depart and ${\rm L}_{\rm{A}}$ may confine whatever is left.  For example, one might
     photoassociate some A atoms into dimers during the initial period, and thereby be left with a nearly
     pure population of dimers trapped in ${\rm L}_{\rm{A}}$ at the end.
     LeBlanc and Thywissen \cite{LeblancThywissen}
     have pointed out the advantage of magic-zero wavelengths for traps containing two species.  If another species,
      B, is added to the model configuration, it will ordinarily be affected by ${\rm L}_{\rm{A}}$,
      so B can be moved by shifting ${\rm L}_{\rm{A}}$, while A remains unaffected.  Schemes of this
      type have been used for entropy transfer and controlled collisions between $^{87}$Rb and $^{41}$K
       \cite{PhysRevLett.103.140401,PhysRevLett.104.153202,newa}.
        For bichromatic optical lattice schemes, such as those discussed
        by Brickman Soderberg, {\em et al.} \cite{Soderberg,klinger:013109},
         it could be useful to incorporate ${\rm L}_{\rm{B}}$ into the model
         configuration, so as to be able to move A and B completely independently.
         In another application, a Sr lattice at a $^3$P$_0$ magic-zero wavelength was
         suggested for realization of quantum information processing \cite{srq}.

 In the next section, we briefly discuss the calculation of frequency-dependent polarizabilities
  of alkali-metal atoms. In section~\ref{sec3}, we
present the magic-zero wavelengths for the alkalis from Li to Cs and discuss some of their
applications.

\section{Frequency-dependent polarizabilities}~\label{sec2}

The background to our approach to calculation of atomic
polarizabilities is treated in a recent review
article~\cite{2010JPhB...43t2001M}.  Here we summarize points salient
to the present work. The frequency-dependent scalar polarizability,
$\alpha^v_0(\omega)$, of an alkali-metal atom in its ground state $
v$ may be separated into a contribution from the core electrons,
$\alpha_{\rm{core}}$, a core modification due to the valence
electron, $\alpha_{vc}$, and a contribution from the valence
electron, $\alpha_v(\omega)$. Since core electrons have excitation
energies in the far-ultraviolet region of the spectrum, the core
polarizability dependends weakly on $\omega$ for the optical
frequencies treated here. Therefore, we approximate the core
polarizability by its dc value as calculated in the random-phase
approximation (RPA) ~\cite{datatab2}, an approach that has been quite
successful in previous applications. The core polarizability is
corrected for Pauli blocking of core-valence excitations by
introducing  an extra term $\alpha_{vc}$. For consistency, this is
also calculated in RPA. Therefore, the ground state polarizability
may be separated as
    \begin{equation}
        \alpha_0(\omega)=\alpha_{\rm{core}} + \alpha_{vc} + \alpha_{0}^v(\omega).
    \end{equation}
The valence contribution to the static ac polarizability is calculated
using the sum-over-states approach~\cite{1}:
     \begin{equation}
         \alpha_{0}^v(\omega)=\frac{2}{3(2j_v+1)}\sum_k\frac{{\left\langle
          k\left\|D\right\|v\right\rangle}^2(E_k-E_v)}{
         (E_k-E_v)^2-\omega^2}, \label{eq-1}
      \end{equation}
where $\left\langle k\left\|D\right\|v\right\rangle$ is the reduced
electric-dipole (E1) matrix element.
 In this equation,
$\omega$ is assumed to be at least several linewidths off resonance with the corresponding transitions.  Unless stated otherwise, we use the conventional system of atomic units, a.u., in which $e, m_{\rm e}$, $4\pi \epsilon_0$ and the reduced Planck
constant $\hbar$ have the numerical value 1.  Polarizability in a.u. has the dimension of volume, and its numerical values presented here are
expressed in units of $a^3_0$, where $a_0\approx0.052918$~nm is the Bohr radius. The atomic units for $\alpha$ can be converted to SI units via
 $\alpha/h$~[Hz/(V/m)$^2$]=2.48832$\times10^{-8}\alpha$~[a.u.], where
 the conversion coefficient is $4\pi \epsilon_0 a^3_0/h$ and the
 Planck constant $h$ is factored out.

\begin{table}
\caption{$5s - np$ contributions to the frequency-dependent
polarizability  of the ground state of Rb at
$\lambda_{\textrm{zero}}=423.0448$~nm = 1/(23638.16 cm$^{-1}$).
Absolute values of electric-dipole matrix elements are expressed in
a.u. ($ea_0$), and the corresponding energy differences are expressed
in conventional wavenumber units (cm$^{-1}$ ). \label{tab1} }
\begin{ruledtabular}
\begin{tabular}{lccr}
\multicolumn{1}{c}{Contribution}& \multicolumn{1}{c}{$|\langle5s\|D\|np_{1/2}\rangle|$}& \multicolumn{1}{c}{$E_{np_{j}}-E_{5s}$}&
\multicolumn{1}{c}{$\alpha_0$}\\
\hline
  $5p_{1/2}$ & 4.231&    12579.0&    -41.130\\
  $6p_{1/2}$ & 0.325&    23715.1&     50.235\\
  $7p_{1/2}$ & 0.115&    27835.0&      0.124\\
  $8p_{1/2}$ & 0.059&    29835.0&      0.023\\
  $np_{1/2}$~tail&  &           &      0.085\\   [0.3pc]
  $5p_{3/2}$ & 5.978 &   12816.5&    -84.938  \\
  $6p_{3/2}$ & 0.528 &   23792.6&     66.140  \\
  $7p_{3/2}$ & 0.202 &   27870.1&      0.383  \\
  $8p_{3/2}$ & 0.111 &   29853.8&      0.081  \\
 $np_{3/2}$~tail&    &               & 0.285      \\    [0.3pc]
 $ \alpha_{\textrm{core}}$&    &               &  9.076  \\
  $\alpha_{vc}$   &     &              & -0.367\\
$\alpha_{0}^v$     &     &              & -8.712\\
  Total $\alpha_{0}(\omega)$    &        &    &  0.00\\
\end{tabular}
\end{ruledtabular}
\end{table}

\begin{figure}
  \includegraphics[width=3.6in]{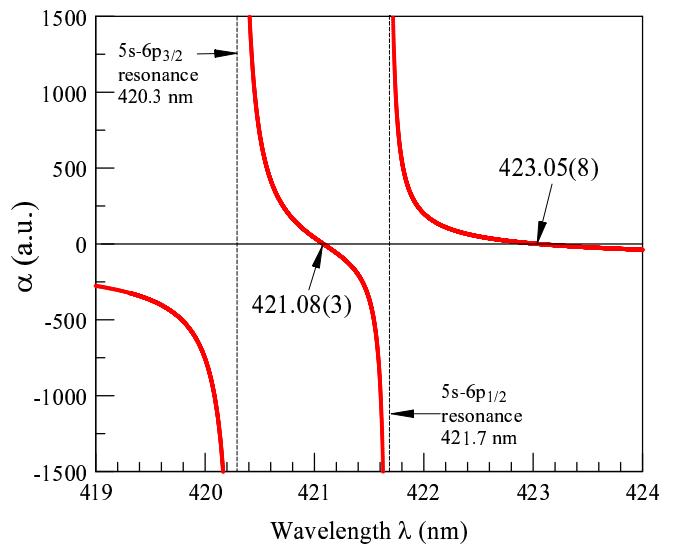}
  \caption{The frequency-dependent polarizability of the Rb ground state.
  The first two zero-magic wavelengths are marked with arrows.}
  \label{fig1}
\end{figure}

\begin{table} [h]
\caption{\label{tab2}  Magic-zero wavelengths $\lambda_{\rm{zero}}$ for alkali-metal atoms from Li to Cs. The resonant wavelengths
$\lambda_{\rm{res}}$ for relevant transitions are also listed. The wavelengths (in vacuum) are given in nm.}
\begin{ruledtabular}
\begin{tabular}{clll}
\multicolumn{1}{l}{Atom}& \multicolumn{1}{l}{Resonance}& \multicolumn{1}{l}{$\lambda_{\rm{res}}$}&
\multicolumn{1}{l}{$\lambda_{\rm{zero}}$}\\
\hline $^6$Li&$2s-2p_{1/2}$     & 670.992478
 &            \\
&                  &       & 670.987445(1)\\
&$2s-2p_{3/2}$     & 670.977380&  \\[0.2pc]
\hline $^7$Li&$2s-2p_{1/2}$     & 670.976658
 &            \\
&                  &       & 670.971626(1)\\
&$2s-2p_{3/2}$     & 670.961561
&  \\
Li&                  &       & 324.18(6)\\
&$2s-3p_{1/2}$ & 323.3576 & \\
&$2s-3p_{3/2}$ & 323.3566 & \\
&                  &       & 274.911(7) \\
&$2s-4p_{1/2}$ & 274.2001& \\[0.2pc]
\hline
Na&$3s-3p_{1/2}$     & 589.7558 &            \\
&                  &       & 589.5565(3) \\
&$3s-3p_{3/2}$     & 589.1583 &  \\
&                  &       & 331.905(3) \\
&$3s-4p_{1/2}$ & 330.3929 & \\
&                  &       &330.3723 \\
&$3s-4p_{3/2}$ & 330.3319 & \\
&                  &       & 285.5817(8) \\
&$3s-5p_{1/2}$ & 285.3850 & \\[0.2pc]
\hline
K&$4s-4p_{1/2}$     & 770.1083 &            \\
  &                &       & 768.971(3) \\
&$4s-4p_{3/2}$     & 766.7009 &  \\
&              &       & 405.98(4) \\
&$4s-5p_{1/2}$ & 404.8356 & \\
&                  &       & 404.72(4) \\
&$4s-5p_{3/2}$ & 404.5285 & \\
&                  &       & 344.933(1) \\
&$4s-6p_{1/2}$ & 344.8363 & \\[0.2pc]
\hline
Rb&$5s-5p_{1/2}$     & 794.9789 &            \\
&                  &       & 790.034(7) \\
&$5s-5p_{3/2}$     & 780.2415 &  \\
&              &       & 423.05(8) \\
&$5s-6p_{1/2}$ & 421.6726 & \\
&                  &       & 421.08(3) \\
&$5s-6p_{3/2}$ & 420.2989 & \\
&                  &       & 359.42(3) \\
&$5s-7p_{1/2}$ & 359.2593 & \\[0.2pc]
\hline
Cs&$6s-6p_{1/2}$     & 894.5929 &            \\
&                  &       &  880.25(4) \\
&$6s-6p_{3/2}$     & 852.3472 &  \\
&              &       & 460.22(2) \\
&$6s-7p_{1/2}$ & 459.4459 & \\
&                  &       & 457.31(3) \\
&$6s-7p_{3/2}$ & 455.6557 & \\
&                  &       & 389.029(4) \\
&$6s-8p_{1/2}$ & 388.9714 &
\end{tabular}
\end{ruledtabular}
\end{table}
The calculation of the ground state frequency-dependent polarizabilities in alkali-metal atoms has been previously discussed in ~\cite{bin1,magic},
and we give only brief summary of the approach.
 The sum  over intermediate $k$ states in Eq.~(\ref{eq-1}) converges rapidly. Therefore, we separate the valence state polarizability into two parts,
  $\alpha_{\rm{main}}$,
containing the contributions from the few lowest $np$ states, and the
remainder, $\alpha_{\rm{tail}}$. We note that our calculations are
carried out with the finite basis set constructed using B-splines
\cite{spline} making the sum finite. In the calculation of
$\alpha_{\rm{main}}$, we use the experimental values compiled in
Ref.~\cite{relsd} along with their uncertainties for the first
$ns-np$ matrix elements, for example the $4s-4p_{j}$ matrix elements
in K. For all other terms, we use the relativistic all-order values
~\cite{review07,relsd} of the matrix elements and the experimental
values of the energies \cite{NIST1,NIST,NIST2}.  In the relativistic
all-order method, all single-double (SD) or single-double and partial
valence triple (SDpT) excitations of the Dirac-Fock (DF) wave
function are included to all orders of perturbation
theory~\cite{relsd,CC2,1}. We conduct additional semi-empirical
scaling of our all-order values (SD$_{\rm{sc}}$) where we expect
scaled values to be more accurate or for more accurate evaluation of
the uncertainties. Our scaling procedure and evaluation of the
uncertainties  of the all-order results have been recently discussed
in Ref.~\cite{ca}. Briefly, the uncertainties of the all-order matrix
elements are given by the spread of their SD, SDpT,  SD$_{\rm{sc}}$
and SDpT$_{\rm{sc}}$ values. These are also used to calculate the
uncertainties in the state-by-state contributions to the
frequency-dependent polarizability. The tail contributions,
$\alpha_{\rm{tail}}$, are calculated in the DF approximation using
complete basis set functions that are linear combinations of
B-splines~\cite{relsdrb}. In the cases treated here, the tail
contribution is of the order of 1\% of the net valence contribution
$\alpha_{0}^v$.

We define the magic-zero  wavelength $\lambda_{\rm{zero}}$ as the wavelength where the
 ac polarizability of the ground state vanishes.  In practice, we calculated $\alpha_0(\omega)$ for a range of values in the vicinity of  relevant resonances and
identified values of $\omega$  where the polarizability turned to zero
with sufficient numerical accuracy.

\begin{table*}
\caption{\label{tab3} Wavelength (in vacuum) of selected transitions in Mg, Ca, Zn, Cd, Sr, Ba, Hg, Yb, Dy, Ho, and Er in nm.}
\begin{ruledtabular}
\begin{tabular}{llclc}
\multicolumn{1}{c}{Atom}& \multicolumn{1}{c}{Transition }& \multicolumn{1}{c}{Wavelength}& \multicolumn{1}{c}{Transition }&
\multicolumn{1}{c}{Wavelength}\\ \hline
 Mg & $3s^2~ ^1S_{0} - 3s3p~ ^1P_1$ & 285.3 & $3s^2\ ^1S_{0} - 3s3p\ ^3P_1$ & 457.2 \\
Ca  & $4s^2~ ^1S_{0} - 4s4p~ ^1P_1$ & 422.8 & $4s^2\ ^1S_{0} - 4s4p\ ^3P_1$ &  657.5\\
Sr  & $5s^2~ ^1S_{0} - 5s5p~ ^1P_1$ & 460.9 & $5s^2\ ^1S_{0} - 5s5p\ ^3P_1$ & 689.5 \\
Ba  & $6s^2~ ^1S_{0} - 6s6p~ ^1P_1$ & 553.7 & $6s^2\ ^1S_{0} - 6s6p\ ^3P_1$ & 791.4 \\
Zn  & $4s^2~ ^1S_{0} - 4s4p~ ^1P_1$ & 213.9 & $4s^2\ ^1S_{0} - 4s4p\ ^3P_1$ &  307.7\\
Cd  & $5s^2~ ^1S_{0} - 5s5p~ ^1P_1$ & 228.9 & $5s^2\ ^1S_{0} - 5s5p\ ^3P_1$ & 326.2 \\
Hg  & $6s^2~ ^1S_{0} - 6s6p~ ^1P_1$ & 184.9 & $6s^2\ ^1S_{0} - 6s6p\ ^3P_1$ & 253.7 \\
Yb  & $6s^2~ ^1S_{0} - 6s6p~ ^1P_1$ & 398.9 & $6s^2\ ^1S_{0} - 6s6p\ ^3P_1$ & 555.8 \\
Dy  & $4f^{10}6s^2\ ^5I_{8} - 4f^{10}(^5I_{8}) ~6s6p~(^1P_1)$ $J=9$ &421.3 & & \\
Er  & $4f^{12}6s^2\ ^3H_{6} - 4f^{11}(^3H_{6}) ~6s6p~ (^3P_1)$ $J=7$ &582.8 & & \\
Ho  & $4f^{11}6s^2\ ^4I_{15/2} - 4f^{11}(^4I_{15/2}) ~6s6p~(^3P_1)$ $J=17/2$ &598.5 & & \\
Ho  & $4f^{11}6s^2\ ^4I_{15/2} - 4f^{11}(^4I_{15/2}) ~6s6p~(^1P_1)$ $J=13/2$ &416.4 & & \\
\end{tabular}
 \end{ruledtabular}
\end{table*}

  We
illustrate the cancellation of all contributions to $5s$ Rb
polarizability at $\lambda_{\rm{zero}}$=423.0448~nm in
Table~\ref{tab1}. Since this wavelength is between $5s-5p_{3/2}$ and
$5s-6p_{1/2}$ resonances, the contributions of the $5p_j$ and $6p_j$
terms strongly dominate. However, the contribution from the core is
significant (11\% of the largest valence term).  This table shows
that $\lambda_{\rm{zero}}$ is located where the valence contribution
to the polarizability cancels the adjusted core contribution, a
feature that is common to all the cases treated here.  The zero
crossing point is in the close vicinity of the $5s-6p_{1/2}$
resonance owing to the relative size of the $5s-5p_j$ and $5s-6p_j$
reduced electric-dipole matrix elements given in the second column of
Table~\ref{tab1}. The $5s-6p$ matrix elements are more than an order
of magnitude smaller than the $5s-5p$ matrix elements. Since
polarizability contributions are proportional to the square of the
matrix element, the denominators of the $6p_j$ terms have to become
very small to cancel out the $5p$ contributions.

This magic-zero wavelength is illustrated in Fig.~\ref{fig1} where we
plot ground-state polarizability of Rb atom in a.u. in the vicinity
of the $5s-6p_j$ resonances. Another zero crossing point shown in the
figure is located between $5s-6p_{1/2}$ and $5s-6p_{3/2}$ resonances,
as expected.  The next magic-zero wavelength will be located close to
the $5s-7p_{1/2}$ resonance since the values of the matrix elements
continue to decrease with $n$.

\section{Results and Applications}~\label{sec3}

In Table~\ref{tab2}, we list the vacuum $\lambda_{\rm{zero}}$ wavelengths for alkali-metal atoms from Li to Cs.  For convenience of presentation, we
also list the resonant wavelength $\lambda_{\rm{res}}$ in vacuum in the relevant range of wavelengths. We order the lists of the resonant wavelengths
and $\lambda_{\rm{zero}}$ to indicate the respective placements of $\lambda_{\rm{zero}}$ and their distances from resonances. The resonant vacuum
wavelength values are obtained from energy levels from National Institute of Standards
 and Technology (NIST) database~\cite{NIST1} with the exception
of the $2s-2p_{1/2}$ and $2s-2p_{3/2}$ transition wavelengths for
$^6$Li and $^7$Li that are taken from recent measurements \cite{Li}.

Since alkali ground states have electric dipole transitions only to
$p$ states, their polarizabilities will cross zero only between two
$ns-n^{\prime}p$ resonances. We set the wavelength of the
$ns-(n+2)p_{1/2}$ resonance as a lower wavelength bound for our
search. The fine structure of the $(n+2)p$ level is sufficiently
small for all alkalis to make the zero point between
$ns-(n+2)p_{1/2}$ and $ns-(n+2)p_{3/2}$ relatively difficult to use
in practice, so we do not list it. We omit the $\lambda_{\rm{zero}}$
between $2s-3p_j$ resonances for the same reason. The wavelengths of
the next zero-crossing near the $ns-(n+3)p_{1/2}$ resonances are in
the ultraviolet, and not as readily accessible in most laboratories,
so we have not calculated them.  However, this would be a routine
matter for future work. There are no $\lambda_{\rm{zero}}$ at
wavelengths greater than those of the primary $ns-np_{1/2}$
resonances. Within these constraints, we have found four
$\lambda_{\rm{zero}}$ for Na, K, Rb, and Cs and three
$\lambda_{\rm{zero}}$ for Li, as shown in Table~\ref{tab1}.

The stated uncertainties in the $\lambda_{\rm{zero}}$ values are
taken to be the maximum difference between the central value and the
crossing of the $\alpha_{{0}}\pm\delta\alpha_{0}$ with zero, where
$\delta\alpha_{0}$ is the uncertainty in the ground state
polarizability value at that wavelength. The uncertainties in the
values of polarizabilities are obtained by adding uncertainties in
the individual polarizability contributions in quadrature.

We find small but significant differences in the first magic
wavelengths of $^6$Li and $^7$Li due to the isotope shift. These
values refer to the centers of gravity of all hyperfine states and do
not take into account the hyperfine structure.  Therefore, this
$\lambda_{\rm{zero}}$ and the corresponding $2s-2p_j$ resonance
wavelengths are listed separately. We verified that isotope shift of
the $2s-2p$ traditions in Li does not affect the next magic
wavelength, 324.18(6)nm, so we use NIST data for the other
transitions.
 We also investigated possible dependence of the first magic-zero wavelengths
 on the isotope shift for $^{39}$K, $^{40}$K, $^{41}$K and $^{85}$Rb, $^{87}$Rb.
  The D1 ($4s-4p_{1/2}$) and D2
($4s-4p_{3/2}$) line wavelengths for $^{39}$K, $^{40}$K, and $^{41}$K have been
measured using a femtosecond laser frequency comb by Falke {\it et
al.} \cite{falke}. We carry out three calculations of the first magic-zero
wavelength using D1, D2 wavelengths for the specific isotope in our
calculations. The resulting value for $^{39}$K and  $^{40}$K,  768.971(3)nm,
is the same as the value quoted in Table~\ref{tab2}.   The $^{41}$K
value is 768.970(3)nm, with the difference being well below our quoted uncertainty.
 The calculations of the first magic wavelength in Rb using D1 and
D2 frequencies  for $^{85}$Rb and  $^{87}$Rb listed in \cite{dline85,dline87}
 gave results identical to result from Table~\ref{tab2}, 790.034(7)nm
that was obtained using NIST data. We note that our values for the
first magic wavelengths are in good agreement with  LeBlanc and
Thywissen \cite{LeblancThywissen} calculations with the exception of
their value  for $^{40}$K.

The first magic-zero wavelength in Rb has been measured to be 789.85(1)nm in \cite{PhysRevLett.103.140401}.
 Some discrepancy with our result is most likely due to approximate linear polarization of the beam in
\cite{PhysRevLett.103.140401}. The difference is compatible with a shift in the magic zero wavelength
 caused by few percent spurious $\sigma^-$
polarization component \cite{com}.

Below, we identify two main applications of magic-zero wavelengths.
First, these wavelengths are advantageous  for  cooling of group-II
and other more complicated atoms,  by sympathetic cooling using an
accompanying   alkali atom.

Recently, group II atoms have been the subject of various experiments
and proposals in atomic clock research and quantum information. BEC
of $^{84}$Sr has been reported recently by two groups
\cite{CWC-1,sr-bec}. The element Yb has four boson and two fermion
isotopes, all of which have been cooled into the microkelvin range.
Several exciting new prospects for quantum information processing
with the ground state nuclear spin have recently been identified in
group II elements \cite{srq}. Also, Sr or Yb are useful for polarized
mixtures of fermions, or Bose-Fermi mixtures, where isotopic mixtures
can be studied.

More complex systems have become of interest in the development of
frequency standards and quantum information processing schemes. For
example, the rare earth holmium is a candidate for quantum
information applications \cite{ho1} due to its rich ground-state
hyperfine structure. Erbium has been a subject of recent experimental
work \cite{er1,er2}, stimulated by
 its possible use in a variety of
applications, including narrow linewidth laser cooling and spectroscopy, unique collision studies, and degenerate bosonic and fermionic gases with
long-range magnetic dipole coupling.

 Some species, particularly fermions,
 are difficult to cool by themselves due to unfavorable ultracold collisional dynamics.
  In such cases, it may be possible to use sympathetic cooling in a mixture of the target
   species and one of the alkalis, where the alkali-metal atom is cooled directly by standard
techniques.  This has recently been demonstrated in Yb:Rb  mixtures
\cite{CWC-10a}. Use of $\lambda_{\rm{zero}}$ trap wavelengths could
allow one to release alkali atoms after the target atoms of the other
species are sufficiently cold, in a hybrid trap configuration that
combines optical and magnetic traps or bichromatic optical traps. If
the final trap configuration utilizes a $\lambda_{\rm{zero}}$
wavelength, strong trapping of the target atom is possible while
alkali atoms will be released by turning off its separate trapping
potential. Since placement of the resonances varies significantly
among the alkali-metal atoms, a wide range of $\lambda_{\rm{zero}}$
is available as shown in Table~\ref{tab3}.

We list the resonant wavelengths  for  variety of atomic systems  in Table~\ref{tab3}.
For consistency with the other tables, we list vacuum
wavelengths obtained from the NIST energy levels database~\cite{NIST2}.
Comparing Tables~\ref{tab2} and \ref{tab3} yields many instances of
resonant transitions that are very close to $\lambda_{\rm{zero}}$. Here are
 few of the very  close cases: Mg~285.3  - Na~285.6, Sr~460.9 - Cs~460.2,
Dy~421.3 - Rb~421.1, Ho~598.5 - Na~589.6.

Magic-zero wavelength laser light may be also useful in three-species cooling
 schemes such as reported in Ref.~\cite{three} by allowing easy release
of one of the species from the trap. The work \cite{three} demonstrated that
 the efficiency of sympathetic cooling of the $^6$Li gas by $^{87}$Rb
was increased by the presence of $^{40}$K through catalytic cooling.

Measurements of the magic-zero wavelengths may be used as
high-precision benchmark tests of theory and determination of the
excited-state matrix elements that are difficult to measure by other
methods. Matrix elements of  $ns-n^{\prime}p$ transitions of
alkali-metal atoms, where $ns$ is the ground state, are difficult to
calculate accurately owing to large correlation corrections and small
values of the final numbers. Experimental measurements of the
$\lambda_{\rm_{zero}}$ predicted in this work will serve an an
excellent benchmark test of the all-order calculations. Moreover, it
will be possible to combine these measurements with theoretical
calculations to infer the values of these small matrix elements. Only
one high-precision measurement of such matrix elements ($6s-7p_j$
transitions in Cs) has been carried out to date ~\cite{beta}.

\section{conclusion}~\label{sec4}
In summary, we calculate magic-zero wavelengths in alkali-metal atoms from
 Li to Cs and estimate their uncertainties. Applications of these magic
wavelengths to sympathetic cooling of group-II and other more
complicated atoms with alkalis are discussed. Special cases where
these wavelengths coincide with strong resonance transition in
group-II atoms, Yb, Dy, Ho, and Er are identified. Measurements of
the magic-zero wavelengths for benchmark tests of theory and
experiment are proposed.

This research was performed under the sponsorship of the US
Department of Commerce, National Institute of Standards and
Technology, and was supported by the National Science Foundation
under Physics Frontiers Center Grant PHY-0822671. This work was
performed in part under the sponsorship of the Department of Science
and Technology, India. We thank Joseph Reader and Craig Sansonetti
for helpful discussions.


\begin{thebibliography}{64}
\expandafter\ifx\csname
natexlab\endcsname\relax\def\natexlab#1{#1}\fi
\expandafter\ifx\csname bibnamefont\endcsname\relax
  \def\bibnamefont#1{#1}\fi
\expandafter\ifx\csname bibfnamefont\endcsname\relax
  \def\bibfnamefont#1{#1}\fi
\expandafter\ifx\csname citenamefont\endcsname\relax
  \def\citenamefont#1{#1}\fi
\expandafter\ifx\csname url\endcsname\relax
  \def\url#1{\texttt{#1}}\fi
\expandafter\ifx\csname urlprefix\endcsname\relax\def\urlprefix{URL
}\fi \providecommand{\bibinfo}[2]{#2}
\providecommand{\eprint}[2][]{\url{#2}}

\bibitem[{\citenamefont{{Truscott} et~al.}(2001)\citenamefont{{Truscott},
  {Strecker}, {McAlexander}, {Partridge}, and {Hulet}}}]{2001Sci...291.2570T}
\bibinfo{author}{\bibfnamefont{A.~G.} \bibnamefont{{Truscott}}},
  \bibinfo{author}{\bibfnamefont{K.~E.} \bibnamefont{{Strecker}}},
  \bibinfo{author}{\bibfnamefont{W.~I.} \bibnamefont{{McAlexander}}},
  \bibinfo{author}{\bibfnamefont{G.~B.} \bibnamefont{{Partridge}}},
  \bibnamefont{and} \bibinfo{author}{\bibfnamefont{R.~G.}
  \bibnamefont{{Hulet}}}, \bibinfo{journal}{Science}
  \textbf{\bibinfo{volume}{291}}, \bibinfo{pages}{2570} (\bibinfo{year}{2001}).

\bibitem[{\citenamefont{Schreck
  et~al.}(2001{\natexlab{a}})\citenamefont{Schreck, Khaykovich, Corwin,
  Ferrari, Bourdel, Cubizolles, and Salomon}}]{PhysRevLett.87.080403}
\bibinfo{author}{\bibfnamefont{F.}~\bibnamefont{Schreck}},
  \bibinfo{author}{\bibfnamefont{L.}~\bibnamefont{Khaykovich}},
  \bibinfo{author}{\bibfnamefont{K.~L.} \bibnamefont{Corwin}},
  \bibinfo{author}{\bibfnamefont{G.}~\bibnamefont{Ferrari}},
  \bibinfo{author}{\bibfnamefont{T.}~\bibnamefont{Bourdel}},
  \bibinfo{author}{\bibfnamefont{J.}~\bibnamefont{Cubizolles}},
  \bibnamefont{and} \bibinfo{author}{\bibfnamefont{C.}~\bibnamefont{Salomon}},
  \bibinfo{journal}{Phys. Rev. Lett.} \textbf{\bibinfo{volume}{87}},
  \bibinfo{pages}{080403} (\bibinfo{year}{2001}{\natexlab{a}}).

\bibitem[{\citenamefont{{Modugno} et~al.}(2001)\citenamefont{{Modugno},
  {Ferrari}, {Roati}, {Brecha}, {Simoni}, and
  {Inguscio}}}]{2001Sci...294.1320M}
\bibinfo{author}{\bibfnamefont{G.}~\bibnamefont{{Modugno}}},
  \bibinfo{author}{\bibfnamefont{G.}~\bibnamefont{{Ferrari}}},
  \bibinfo{author}{\bibfnamefont{G.}~\bibnamefont{{Roati}}},
  \bibinfo{author}{\bibfnamefont{R.~J.} \bibnamefont{{Brecha}}},
  \bibinfo{author}{\bibfnamefont{A.}~\bibnamefont{{Simoni}}}, \bibnamefont{and}
  \bibinfo{author}{\bibfnamefont{M.}~\bibnamefont{{Inguscio}}},
  \bibinfo{journal}{Science} \textbf{\bibinfo{volume}{294}},
  \bibinfo{pages}{1320} (\bibinfo{year}{2001}).

\bibitem[{\citenamefont{{Ivanov} et~al.}(2011)\citenamefont{{Ivanov},
  {Khramov}, {Hansen}, {Dowd}, {Muenchow}, {Jamison}, and {Gupta}}}]{new1}
\bibinfo{author}{\bibfnamefont{V.~V.} \bibnamefont{{Ivanov}}},
  \bibinfo{author}{\bibfnamefont{A.}~\bibnamefont{{Khramov}}},
  \bibinfo{author}{\bibfnamefont{A.~H.} \bibnamefont{{Hansen}}},
  \bibinfo{author}{\bibfnamefont{W.~H.} \bibnamefont{{Dowd}}},
  \bibinfo{author}{\bibfnamefont{F.}~\bibnamefont{{Muenchow}}},
  \bibinfo{author}{\bibfnamefont{A.~O.} \bibnamefont{{Jamison}}},
  \bibnamefont{and} \bibinfo{author}{\bibfnamefont{S.}~\bibnamefont{{Gupta}}},
  \bibinfo{journal}{Phys. Rev. Lett.} \textbf{\bibinfo{volume}{106}},
  \bibinfo{pages}{153201} (\bibinfo{year}{2011}).

\bibitem[{\citenamefont{{Baumer} et~al.}(2011)\citenamefont{{Baumer},
  {M{\"u}nchow}, {G{\"o}rlitz}, {Maxwell}, {Julienne}, and {Tiesinga}}}]{new2}
\bibinfo{author}{\bibfnamefont{F.}~\bibnamefont{{Baumer}}},
  \bibinfo{author}{\bibfnamefont{F.}~\bibnamefont{{M{\"u}nchow}}},
  \bibinfo{author}{\bibfnamefont{A.}~\bibnamefont{{G{\"o}rlitz}}},
  \bibinfo{author}{\bibfnamefont{S.~E.} \bibnamefont{{Maxwell}}},
  \bibinfo{author}{\bibfnamefont{P.~S.} \bibnamefont{{Julienne}}},
  \bibnamefont{and}
  \bibinfo{author}{\bibfnamefont{E.}~\bibnamefont{{Tiesinga}}},
  \bibinfo{journal}{ArXiv e-prints}  (\bibinfo{year}{2011}),
  \eprint{1104.1722}.

\bibitem[{\citenamefont{Papp and Wieman}(2006)}]{PhysRevLett.97.180404}
\bibinfo{author}{\bibfnamefont{S.~B.} \bibnamefont{Papp}} \bibnamefont{and}
  \bibinfo{author}{\bibfnamefont{C.~E.} \bibnamefont{Wieman}},
  \bibinfo{journal}{Phys. Rev. Lett.} \textbf{\bibinfo{volume}{97}},
  \bibinfo{pages}{180404} (\bibinfo{year}{2006}).

\bibitem[{\citenamefont{Ospelkaus et~al.}(2006)\citenamefont{Ospelkaus,
  Ospelkaus, Humbert, Ernst, Sengstock, and Bongs}}]{PhysRevLett.97.120402}
\bibinfo{author}{\bibfnamefont{C.}~\bibnamefont{Ospelkaus}},
  \bibinfo{author}{\bibfnamefont{S.}~\bibnamefont{Ospelkaus}},
  \bibinfo{author}{\bibfnamefont{L.}~\bibnamefont{Humbert}},
  \bibinfo{author}{\bibfnamefont{P.}~\bibnamefont{Ernst}},
  \bibinfo{author}{\bibfnamefont{K.}~\bibnamefont{Sengstock}},
  \bibnamefont{and} \bibinfo{author}{\bibfnamefont{K.}~\bibnamefont{Bongs}},
  \bibinfo{journal}{Phys. Rev. Lett.} \textbf{\bibinfo{volume}{97}},
  \bibinfo{pages}{120402} (\bibinfo{year}{2006}).

\bibitem[{\citenamefont{{Ospelkaus} and
  {Ospelkaus}}(2008)}]{2008JPhB...41t3001O}
\bibinfo{author}{\bibfnamefont{C.}~\bibnamefont{{Ospelkaus}}} \bibnamefont{and}
  \bibinfo{author}{\bibfnamefont{S.}~\bibnamefont{{Ospelkaus}}},
  \bibinfo{journal}{J. Phys. B} \textbf{\bibinfo{volume}{41}},
  \bibinfo{pages}{203001} (\bibinfo{year}{2008}).

\bibitem[{\citenamefont{Ni et~al.}(2008)\citenamefont{Ni, Ospelkaus,
  de~Miranda, Pe'er, Neyenhuis, Zirbel, Kotochigova, Julienne, Jin, and
  Ye}}]{Ni}
\bibinfo{author}{\bibfnamefont{K.-K.} \bibnamefont{Ni}},
  \bibinfo{author}{\bibfnamefont{S.}~\bibnamefont{Ospelkaus}},
  \bibinfo{author}{\bibfnamefont{M.~H.~G.} \bibnamefont{de~Miranda}},
  \bibinfo{author}{\bibfnamefont{A.}~\bibnamefont{Pe'er}},
  \bibinfo{author}{\bibfnamefont{B.}~\bibnamefont{Neyenhuis}},
  \bibinfo{author}{\bibfnamefont{J.~J.} \bibnamefont{Zirbel}},
  \bibinfo{author}{\bibfnamefont{S.}~\bibnamefont{Kotochigova}},
  \bibinfo{author}{\bibfnamefont{P.~S.} \bibnamefont{Julienne}},
  \bibinfo{author}{\bibfnamefont{D.~S.} \bibnamefont{Jin}}, \bibnamefont{and}
  \bibinfo{author}{\bibfnamefont{J.}~\bibnamefont{Ye}},
  \bibinfo{journal}{Science} \textbf{\bibinfo{volume}{322}},
  \bibinfo{pages}{231} (\bibinfo{year}{2008}).

\bibitem[{\citenamefont{Voigt et~al.}(2009)\citenamefont{Voigt, Taglieber,
  Costa, Aoki, Wieser, H\"ansch, and Dieckmann}}]{PhysRevLett.102.020405}
\bibinfo{author}{\bibfnamefont{A.-C.} \bibnamefont{Voigt}},
  \bibinfo{author}{\bibfnamefont{M.}~\bibnamefont{Taglieber}},
  \bibinfo{author}{\bibfnamefont{L.}~\bibnamefont{Costa}},
  \bibinfo{author}{\bibfnamefont{T.}~\bibnamefont{Aoki}},
  \bibinfo{author}{\bibfnamefont{W.}~\bibnamefont{Wieser}},
  \bibinfo{author}{\bibfnamefont{T.~W.} \bibnamefont{H\"ansch}},
  \bibnamefont{and}
  \bibinfo{author}{\bibfnamefont{K.}~\bibnamefont{Dieckmann}},
  \bibinfo{journal}{Phys. Rev. Lett.} \textbf{\bibinfo{volume}{102}},
  \bibinfo{pages}{020405} (\bibinfo{year}{2009}).

\bibitem[{\citenamefont{Ospelkaus et~al.}(2010)\citenamefont{Ospelkaus, Ni,
  Wang, de~Miranda, Neyenhuis, Quemener, Julienne, Bohn, Jin, and
  Ye}}]{qstate2010}
\bibinfo{author}{\bibfnamefont{S.}~\bibnamefont{Ospelkaus}},
  \bibinfo{author}{\bibfnamefont{K.-K.} \bibnamefont{Ni}},
  \bibinfo{author}{\bibfnamefont{D.}~\bibnamefont{Wang}},
  \bibinfo{author}{\bibfnamefont{M.~H.~G.} \bibnamefont{de~Miranda}},
  \bibinfo{author}{\bibfnamefont{B.}~\bibnamefont{Neyenhuis}},
  \bibinfo{author}{\bibfnamefont{G.}~\bibnamefont{Quemener}},
  \bibinfo{author}{\bibfnamefont{P.~S.} \bibnamefont{Julienne}},
  \bibinfo{author}{\bibfnamefont{J.~L.} \bibnamefont{Bohn}},
  \bibinfo{author}{\bibfnamefont{D.~S.} \bibnamefont{Jin}}, \bibnamefont{and}
  \bibinfo{author}{\bibfnamefont{J.}~\bibnamefont{Ye}},
  \bibinfo{journal}{Science} \textbf{\bibinfo{volume}{327}},
  \bibinfo{pages}{853} (\bibinfo{year}{2010}).

\bibitem[{\citenamefont{DeMille}(2002)}]{mille}
\bibinfo{author}{\bibfnamefont{D.}~\bibnamefont{DeMille}},
  \bibinfo{journal}{Phys. Rev. Lett.} \textbf{\bibinfo{volume}{88}},
  \bibinfo{pages}{067901} (\bibinfo{year}{2002}).

\bibitem[{\citenamefont{Micheli et~al.}(2006)\citenamefont{Micheli, Brennen,
  and Zoller}}]{micheli}
\bibinfo{author}{\bibfnamefont{A.}~\bibnamefont{Micheli}},
  \bibinfo{author}{\bibfnamefont{G.~K.} \bibnamefont{Brennen}},
  \bibnamefont{and} \bibinfo{author}{\bibfnamefont{P.}~\bibnamefont{Zoller}},
  \bibinfo{journal}{Nature Physics} \textbf{\bibinfo{volume}{2}},
  \bibinfo{pages}{341} (\bibinfo{year}{2006}).

\bibitem[{\citenamefont{{Carr} et~al.}(2009)\citenamefont{{Carr}, {DeMille},
  {Krems}, and {Ye}}}]{2009NJPh...11e5049C}
\bibinfo{author}{\bibfnamefont{L.~D.} \bibnamefont{{Carr}}},
  \bibinfo{author}{\bibfnamefont{D.}~\bibnamefont{{DeMille}}},
  \bibinfo{author}{\bibfnamefont{R.~V.} \bibnamefont{{Krems}}},
  \bibnamefont{and} \bibinfo{author}{\bibfnamefont{J.}~\bibnamefont{{Ye}}},
  \bibinfo{journal}{New Journal of Physics} \textbf{\bibinfo{volume}{11}},
  \bibinfo{pages}{055049} (\bibinfo{year}{2009}).

\bibitem[{\citenamefont{Hudson et~al.}(2002)\citenamefont{Hudson, Sauer,
  Tarbutt, and Hinds}}]{hudson}
\bibinfo{author}{\bibfnamefont{J.~J.} \bibnamefont{Hudson}},
  \bibinfo{author}{\bibfnamefont{B.~E.} \bibnamefont{Sauer}},
  \bibinfo{author}{\bibfnamefont{M.~R.} \bibnamefont{Tarbutt}},
  \bibnamefont{and} \bibinfo{author}{\bibfnamefont{E.~A.} \bibnamefont{Hinds}},
  \bibinfo{journal}{Phys. Rev. Lett.} \textbf{\bibinfo{volume}{89}},
  \bibinfo{pages}{023003} (\bibinfo{year}{2002}).

\bibitem[{\citenamefont{Shafer-Ray}(2006)}]{ray}
\bibinfo{author}{\bibfnamefont{N.~E.} \bibnamefont{Shafer-Ray}},
  \bibinfo{journal}{Rhys. Rev. A} \textbf{\bibinfo{volume}{73}},
  \bibinfo{pages}{034102} (\bibinfo{year}{2006}).

\bibitem[{\citenamefont{Meyer and Bohn}(2009)}]{PhysRevA.80.042508}
\bibinfo{author}{\bibfnamefont{E.~R.} \bibnamefont{Meyer}} \bibnamefont{and}
  \bibinfo{author}{\bibfnamefont{J.~L.} \bibnamefont{Bohn}},
  \bibinfo{journal}{Phys. Rev. A} \textbf{\bibinfo{volume}{80}},
  \bibinfo{pages}{042508} (\bibinfo{year}{2009}).

\bibitem[{\citenamefont{{Tassy} et~al.}(2010)\citenamefont{{Tassy}, {Nemitz},
  {Baumer}, {H{\"o}hl}, {Bat{\"a}r}, and {G{\"o}rlitz}}}]{tassy}
\bibinfo{author}{\bibfnamefont{S.}~\bibnamefont{{Tassy}}},
  \bibinfo{author}{\bibfnamefont{N.}~\bibnamefont{{Nemitz}}},
  \bibinfo{author}{\bibfnamefont{F.}~\bibnamefont{{Baumer}}},
  \bibinfo{author}{\bibfnamefont{C.}~\bibnamefont{{H{\"o}hl}}},
  \bibinfo{author}{\bibfnamefont{A.}~\bibnamefont{{Bat{\"a}r}}},
  \bibnamefont{and}
  \bibinfo{author}{\bibfnamefont{A.}~\bibnamefont{{G{\"o}rlitz}}},
  \bibinfo{journal}{J. Phys. B} \textbf{\bibinfo{volume}{43}},
  \bibinfo{pages}{205309} (\bibinfo{year}{2010}).

\bibitem[{\citenamefont{Hadzibabic
  et~al.}(2002{\natexlab{a}})\citenamefont{Hadzibabic, Stan, Dieckmann, Gupta,
  Zwierlein, G\"{o}rlitz, and Ketterle}}]{hadzi}
\bibinfo{author}{\bibfnamefont{Z.}~\bibnamefont{Hadzibabic}},
  \bibinfo{author}{\bibfnamefont{C.~A.} \bibnamefont{Stan}},
  \bibinfo{author}{\bibfnamefont{K.}~\bibnamefont{Dieckmann}},
  \bibinfo{author}{\bibfnamefont{S.}~\bibnamefont{Gupta}},
  \bibinfo{author}{\bibfnamefont{M.~W.} \bibnamefont{Zwierlein}},
  \bibinfo{author}{\bibfnamefont{A.}~\bibnamefont{G\"{o}rlitz}},
  \bibnamefont{and} \bibinfo{author}{\bibfnamefont{W.}~\bibnamefont{Ketterle}},
  \bibinfo{journal}{Phys. Rev. Lett.} \textbf{\bibinfo{volume}{88}},
  \bibinfo{pages}{160401} (\bibinfo{year}{2002}{\natexlab{a}}).

\bibitem[{\citenamefont{Roati et~al.}(2002)\citenamefont{Roati, Riboli,
  Modugno, and Inguscio}}]{Roati}
\bibinfo{author}{\bibfnamefont{G.}~\bibnamefont{Roati}},
  \bibinfo{author}{\bibfnamefont{F.}~\bibnamefont{Riboli}},
  \bibinfo{author}{\bibfnamefont{G.}~\bibnamefont{Modugno}}, \bibnamefont{and}
  \bibinfo{author}{\bibfnamefont{M.}~\bibnamefont{Inguscio}},
  \bibinfo{journal}{Phys. Rev. Lett.} \textbf{\bibinfo{volume}{89}},
  \bibinfo{pages}{150403} (\bibinfo{year}{2002}).

\bibitem[{\citenamefont{Regal et~al.}(2004)\citenamefont{Regal, Greiner, and
  Jin}}]{regal2}
\bibinfo{author}{\bibfnamefont{C.~A.} \bibnamefont{Regal}},
  \bibinfo{author}{\bibfnamefont{M.}~\bibnamefont{Greiner}}, \bibnamefont{and}
  \bibinfo{author}{\bibfnamefont{D.~S.} \bibnamefont{Jin}},
  \bibinfo{journal}{Phys. Rev. Lett.} \textbf{\bibinfo{volume}{92}},
  \bibinfo{pages}{040403} (\bibinfo{year}{2004}).

\bibitem[{\citenamefont{Chin et~al.}(2004)\citenamefont{Chin, Bartenstein,
  Altmeyer, Riedl, Jochim, Denschlag, and Grimm}}]{chin}
\bibinfo{author}{\bibfnamefont{C.}~\bibnamefont{Chin}},
  \bibinfo{author}{\bibfnamefont{M.}~\bibnamefont{Bartenstein}},
  \bibinfo{author}{\bibfnamefont{A.}~\bibnamefont{Altmeyer}},
  \bibinfo{author}{\bibfnamefont{S.}~\bibnamefont{Riedl}},
  \bibinfo{author}{\bibfnamefont{S.}~\bibnamefont{Jochim}},
  \bibinfo{author}{\bibfnamefont{J.~H.} \bibnamefont{Denschlag}},
  \bibnamefont{and} \bibinfo{author}{\bibfnamefont{R.}~\bibnamefont{Grimm}},
  \bibinfo{journal}{Science} \textbf{\bibinfo{volume}{305}},
  \bibinfo{pages}{1128} (\bibinfo{year}{2004}).

\bibitem[{\citenamefont{Kinast et~al.}(2004)\citenamefont{Kinast, Hemmer, Gehm,
  Turlapov, and Thomas}}]{kinast}
\bibinfo{author}{\bibfnamefont{J.}~\bibnamefont{Kinast}},
  \bibinfo{author}{\bibfnamefont{S.~L.} \bibnamefont{Hemmer}},
  \bibinfo{author}{\bibfnamefont{M.~E.} \bibnamefont{Gehm}},
  \bibinfo{author}{\bibfnamefont{A.}~\bibnamefont{Turlapov}}, \bibnamefont{and}
  \bibinfo{author}{\bibfnamefont{J.~E.} \bibnamefont{Thomas}},
  \bibinfo{journal}{Phys. Rev. Lett.} \textbf{\bibinfo{volume}{92}},
  \bibinfo{pages}{150402} (\bibinfo{year}{2004}).

\bibitem[{\citenamefont{{Stan} et~al.}(2004)\citenamefont{{Stan}, {Zwierlein},
  {Schunck}, {Raupach}, and {Ketterle}}}]{Stan}
\bibinfo{author}{\bibfnamefont{C.~A.} \bibnamefont{{Stan}}},
  \bibinfo{author}{\bibfnamefont{M.~W.} \bibnamefont{{Zwierlein}}},
  \bibinfo{author}{\bibfnamefont{C.~H.} \bibnamefont{{Schunck}}},
  \bibinfo{author}{\bibfnamefont{S.~M.} \bibnamefont{{Raupach}}},
  \bibnamefont{and}
  \bibinfo{author}{\bibfnamefont{W.}~\bibnamefont{{Ketterle}}},
  \bibinfo{journal}{Phys. Rev. Lett.} \textbf{\bibinfo{volume}{93}},
  \bibinfo{pages}{143001} (\bibinfo{year}{2004}).

\bibitem[{\citenamefont{{Inouye} et~al.}(2004)\citenamefont{{Inouye},
  {Goldwin}, {Olsen}, {Ticknor}, {Bohn}, and {Jin}}}]{Inouye}
\bibinfo{author}{\bibfnamefont{S.}~\bibnamefont{{Inouye}}},
  \bibinfo{author}{\bibfnamefont{J.}~\bibnamefont{{Goldwin}}},
  \bibinfo{author}{\bibfnamefont{M.~L.} \bibnamefont{{Olsen}}},
  \bibinfo{author}{\bibfnamefont{C.}~\bibnamefont{{Ticknor}}},
  \bibinfo{author}{\bibfnamefont{J.~L.} \bibnamefont{{Bohn}}},
  \bibnamefont{and} \bibinfo{author}{\bibfnamefont{D.~S.} \bibnamefont{{Jin}}},
  \bibinfo{journal}{Phys. Rev. Lett.} \textbf{\bibinfo{volume}{93}},
  \bibinfo{pages}{183201} (\bibinfo{year}{2004}).

\bibitem[{\citenamefont{Naik et~al.}(2011)\citenamefont{Naik, Trenkwalder,
  Kohstall, Spiegelhalder, Zaccanti, Hendl, Schreck, Grimm, Hanna, and
  Julienne}}]{springerlink:10.1140/epjd/e2010-10591-2}
\bibinfo{author}{\bibfnamefont{D.}~\bibnamefont{Naik}},
  \bibinfo{author}{\bibfnamefont{A.}~\bibnamefont{Trenkwalder}},
  \bibinfo{author}{\bibfnamefont{C.}~\bibnamefont{Kohstall}},
  \bibinfo{author}{\bibfnamefont{F.~M.} \bibnamefont{Spiegelhalder}},
  \bibinfo{author}{\bibfnamefont{M.}~\bibnamefont{Zaccanti}},
  \bibinfo{author}{\bibfnamefont{G.}~\bibnamefont{Hendl}},
  \bibinfo{author}{\bibfnamefont{F.}~\bibnamefont{Schreck}},
  \bibinfo{author}{\bibfnamefont{R.}~\bibnamefont{Grimm}},
  \bibinfo{author}{\bibfnamefont{T.~M.} \bibnamefont{Hanna}}, \bibnamefont{and}
  \bibinfo{author}{\bibfnamefont{P.~S.} \bibnamefont{Julienne}}
  (\bibinfo{year}{2011}), \bibinfo{note}{{E}ur. Phys. J. D, DOI:
  10.1140/epjd/e2010-10591-2}.

\bibitem[{\citenamefont{Schreck
  et~al.}(2001{\natexlab{b}})\citenamefont{Schreck, Khaykovich, Corwin,
  Ferrari, Bourdel, Cubizolles, and Salomon}}]{schreck:080403}
\bibinfo{author}{\bibfnamefont{F.}~\bibnamefont{Schreck}},
  \bibinfo{author}{\bibfnamefont{L.}~\bibnamefont{Khaykovich}},
  \bibinfo{author}{\bibfnamefont{K.~L.} \bibnamefont{Corwin}},
  \bibinfo{author}{\bibfnamefont{G.}~\bibnamefont{Ferrari}},
  \bibinfo{author}{\bibfnamefont{T.}~\bibnamefont{Bourdel}},
  \bibinfo{author}{\bibfnamefont{J.}~\bibnamefont{Cubizolles}},
  \bibnamefont{and} \bibinfo{author}{\bibfnamefont{C.}~\bibnamefont{Salomon}},
  \bibinfo{journal}{Phys.\ Rev.\ Lett.} \textbf{\bibinfo{volume}{87}},
  \bibinfo{eid}{080403} (\bibinfo{year}{2001}{\natexlab{b}}).

\bibitem[{\citenamefont{Hadzibabic
  et~al.}(2002{\natexlab{b}})\citenamefont{Hadzibabic, Stan, Dieckmann, Gupta,
  Zwierlein, G\"{o}rlitz, and Ketterle}}]{hadzibabic:160401}
\bibinfo{author}{\bibfnamefont{Z.}~\bibnamefont{Hadzibabic}},
  \bibinfo{author}{\bibfnamefont{C.~A.} \bibnamefont{Stan}},
  \bibinfo{author}{\bibfnamefont{K.}~\bibnamefont{Dieckmann}},
  \bibinfo{author}{\bibfnamefont{S.}~\bibnamefont{Gupta}},
  \bibinfo{author}{\bibfnamefont{M.~W.} \bibnamefont{Zwierlein}},
  \bibinfo{author}{\bibfnamefont{A.}~\bibnamefont{G\"{o}rlitz}},
  \bibnamefont{and} \bibinfo{author}{\bibfnamefont{W.}~\bibnamefont{Ketterle}},
  \bibinfo{journal}{Phys.\ Rev.\ Lett.} \textbf{\bibinfo{volume}{88}},
  \bibinfo{eid}{160401} (\bibinfo{year}{2002}{\natexlab{b}}).

\bibitem[{\citenamefont{Modugno et~al.}(2002)\citenamefont{Modugno, Roati,
  Riboli, Ferlaino, Brecha, and Inguscio}}]{Modugno}
\bibinfo{author}{\bibfnamefont{G.}~\bibnamefont{Modugno}},
  \bibinfo{author}{\bibfnamefont{G.}~\bibnamefont{Roati}},
  \bibinfo{author}{\bibfnamefont{F.}~\bibnamefont{Riboli}},
  \bibinfo{author}{\bibfnamefont{F.}~\bibnamefont{Ferlaino}},
  \bibinfo{author}{\bibfnamefont{R.~J.} \bibnamefont{Brecha}},
  \bibnamefont{and} \bibinfo{author}{\bibfnamefont{M.}~\bibnamefont{Inguscio}},
  \bibinfo{journal}{Science} \textbf{\bibinfo{volume}{297}},
  \bibinfo{pages}{2240} (\bibinfo{year}{2002}).

\bibitem[{\citenamefont{Grimm et~al.}(2000)\citenamefont{Grimm, Weidemuller,
  and Ovchinnikov}}]{grimm}
\bibinfo{author}{\bibfnamefont{R.}~\bibnamefont{Grimm}},
  \bibinfo{author}{\bibfnamefont{M.}~\bibnamefont{Weidemuller}},
  \bibnamefont{and} \bibinfo{author}{\bibfnamefont{Y.~B.}
  \bibnamefont{Ovchinnikov}}, \bibinfo{journal}{Adv. At. Mol. Opt. Phys.}
  \textbf{\bibinfo{volume}{42}} (\bibinfo{year}{2000}).

\bibitem[{\citenamefont{{Leblanc} and {Thywissen}}(2007)}]{LeblancThywissen}
\bibinfo{author}{\bibfnamefont{L.~J.} \bibnamefont{{Leblanc}}}
  \bibnamefont{and} \bibinfo{author}{\bibfnamefont{J.~H.}
  \bibnamefont{{Thywissen}}}, \bibinfo{journal}{Phys. Rev. A}
  \textbf{\bibinfo{volume}{75}}, \bibinfo{pages}{053612}
  (\bibinfo{year}{2007}).

\bibitem[{\citenamefont{Catani et~al.}(2009)\citenamefont{Catani, Barontini,
  Lamporesi, Rabatti, Thalhammer, Minardi, Stringari, and
  Inguscio}}]{PhysRevLett.103.140401}
\bibinfo{author}{\bibfnamefont{J.}~\bibnamefont{Catani}},
  \bibinfo{author}{\bibfnamefont{G.}~\bibnamefont{Barontini}},
  \bibinfo{author}{\bibfnamefont{G.}~\bibnamefont{Lamporesi}},
  \bibinfo{author}{\bibfnamefont{F.}~\bibnamefont{Rabatti}},
  \bibinfo{author}{\bibfnamefont{G.}~\bibnamefont{Thalhammer}},
  \bibinfo{author}{\bibfnamefont{F.}~\bibnamefont{Minardi}},
  \bibinfo{author}{\bibfnamefont{S.}~\bibnamefont{Stringari}},
  \bibnamefont{and} \bibinfo{author}{\bibfnamefont{M.}~\bibnamefont{Inguscio}},
  \bibinfo{journal}{Phys. Rev. Lett.} \textbf{\bibinfo{volume}{103}},
  \bibinfo{pages}{140401} (\bibinfo{year}{2009}).

\bibitem[{\citenamefont{Lamporesi et~al.}(2010)\citenamefont{Lamporesi, Catani,
  Barontini, Nishida, Inguscio, and Minardi}}]{PhysRevLett.104.153202}
\bibinfo{author}{\bibfnamefont{G.}~\bibnamefont{Lamporesi}},
  \bibinfo{author}{\bibfnamefont{J.}~\bibnamefont{Catani}},
  \bibinfo{author}{\bibfnamefont{G.}~\bibnamefont{Barontini}},
  \bibinfo{author}{\bibfnamefont{Y.}~\bibnamefont{Nishida}},
  \bibinfo{author}{\bibfnamefont{M.}~\bibnamefont{Inguscio}}, \bibnamefont{and}
  \bibinfo{author}{\bibfnamefont{F.}~\bibnamefont{Minardi}},
  \bibinfo{journal}{Phys. Rev. Lett.} \textbf{\bibinfo{volume}{104}},
  \bibinfo{pages}{153202} (\bibinfo{year}{2010}).

\bibitem[{\citenamefont{McKay and DeMarco}(2011)}]{newa}
\bibinfo{author}{\bibfnamefont{D.~C.} \bibnamefont{McKay}} \bibnamefont{and}
  \bibinfo{author}{\bibfnamefont{B.}~\bibnamefont{DeMarco}},
  \bibinfo{journal}{Rep. Prog. Phys.} \textbf{\bibinfo{volume}{74}},
  \bibinfo{pages}{054401} (\bibinfo{year}{2011}).

\bibitem[{\citenamefont{{Brickman Soderberg}
  et~al.}(2009)\citenamefont{{Brickman Soderberg}, {Gemelke}, and
  {Chin}}}]{Soderberg}
\bibinfo{author}{\bibfnamefont{K.}~\bibnamefont{{Brickman Soderberg}}},
  \bibinfo{author}{\bibfnamefont{N.}~\bibnamefont{{Gemelke}}},
  \bibnamefont{and} \bibinfo{author}{\bibfnamefont{C.}~\bibnamefont{{Chin}}},
  \bibinfo{journal}{New J. Phys.} \textbf{\bibinfo{volume}{11}},
  \bibinfo{pages}{055022} (\bibinfo{year}{2009}).

\bibitem[{\citenamefont{Klinger et~al.}(2010)\citenamefont{Klinger, Degenkolb,
  Gemelke, Soderberg, and Chin}}]{klinger:013109}
\bibinfo{author}{\bibfnamefont{A.}~\bibnamefont{Klinger}},
  \bibinfo{author}{\bibfnamefont{S.}~\bibnamefont{Degenkolb}},
  \bibinfo{author}{\bibfnamefont{N.}~\bibnamefont{Gemelke}},
  \bibinfo{author}{\bibfnamefont{K.-A.~B.} \bibnamefont{Soderberg}},
  \bibnamefont{and} \bibinfo{author}{\bibfnamefont{C.}~\bibnamefont{Chin}},
  \bibinfo{journal}{Review of Scientific Instruments}
  \textbf{\bibinfo{volume}{81}}, \bibinfo{eid}{013109} (\bibinfo{year}{2010}).

\bibitem[{\citenamefont{Daley et~al.}(2008)\citenamefont{Daley, Boyd, Ye, and
  Zoller}}]{srq}
\bibinfo{author}{\bibfnamefont{A.~J.} \bibnamefont{Daley}},
  \bibinfo{author}{\bibfnamefont{M.~M.} \bibnamefont{Boyd}},
  \bibinfo{author}{\bibfnamefont{J.}~\bibnamefont{Ye}}, \bibnamefont{and}
  \bibinfo{author}{\bibfnamefont{P.}~\bibnamefont{Zoller}},
  \bibinfo{journal}{Phys. Rev. Lett.} \textbf{\bibinfo{volume}{101}},
  \bibinfo{pages}{170504} (\bibinfo{year}{2008}).

\bibitem[{\citenamefont{{Mitroy} et~al.}(2010)\citenamefont{{Mitroy},
  {Safronova}, and {Clark}}}]{2010JPhB...43t2001M}
\bibinfo{author}{\bibfnamefont{J.}~\bibnamefont{{Mitroy}}},
  \bibinfo{author}{\bibfnamefont{M.~S.} \bibnamefont{{Safronova}}},
  \bibnamefont{and} \bibinfo{author}{\bibfnamefont{C.~W.}
  \bibnamefont{{Clark}}}, \bibinfo{journal}{J. Phys. B}
  \textbf{\bibinfo{volume}{43}}, \bibinfo{pages}{202001}
  (\bibinfo{year}{2010}).

\bibitem[{\citenamefont{Johnson et~al.}(1983)\citenamefont{Johnson, Kolb, and
  Huang}}]{datatab2}
\bibinfo{author}{\bibfnamefont{W.~R.} \bibnamefont{Johnson}},
  \bibinfo{author}{\bibfnamefont{D.}~\bibnamefont{Kolb}}, \bibnamefont{and}
  \bibinfo{author}{\bibfnamefont{K.-N.} \bibnamefont{Huang}},
  \bibinfo{journal}{At. Data Nucl. Data Tables} \textbf{\bibinfo{volume}{28}},
  \bibinfo{pages}{334} (\bibinfo{year}{1983}).

\bibitem[{\citenamefont{Safronova and Clark}(2004)}]{1}
\bibinfo{author}{\bibfnamefont{M.~S.} \bibnamefont{Safronova}}
  \bibnamefont{and} \bibinfo{author}{\bibfnamefont{C.~W.} \bibnamefont{Clark}},
  \bibinfo{journal}{Phys. Rev. A} \textbf{\bibinfo{volume}{69}},
  \bibinfo{pages}{040501(R)} (\bibinfo{year}{2004}).

\bibitem[{\citenamefont{Safronova et~al.}(2006)\citenamefont{Safronova, Arora,
  and Clark}}]{bin1}
\bibinfo{author}{\bibfnamefont{M.~S.} \bibnamefont{Safronova}},
  \bibinfo{author}{\bibfnamefont{B.}~\bibnamefont{Arora}}, \bibnamefont{and}
  \bibinfo{author}{\bibfnamefont{C.~W.} \bibnamefont{Clark}},
  \bibinfo{journal}{Phys. Rev. A} \textbf{\bibinfo{volume}{73}},
  \bibinfo{pages}{022505} (\bibinfo{year}{2006}).

\bibitem[{\citenamefont{Arora et~al.}(2007)\citenamefont{Arora, Safronova, and
  Clark}}]{magic}
\bibinfo{author}{\bibfnamefont{B.}~\bibnamefont{Arora}},
  \bibinfo{author}{\bibfnamefont{M.~S.} \bibnamefont{Safronova}},
  \bibnamefont{and} \bibinfo{author}{\bibfnamefont{C.~W.} \bibnamefont{Clark}},
  \bibinfo{journal}{Phys. Rev. A.} \textbf{\bibinfo{volume}{76}},
  \bibinfo{pages}{052509} (\bibinfo{year}{2007}).

\bibitem[{\citenamefont{{Johnson} et~al.}(1988)\citenamefont{{Johnson},
  {Blundell}, and {Sapirstein}}}]{spline}
\bibinfo{author}{\bibfnamefont{W.~R.} \bibnamefont{{Johnson}}},
  \bibinfo{author}{\bibfnamefont{S.~A.} \bibnamefont{{Blundell}}},
  \bibnamefont{and}
  \bibinfo{author}{\bibfnamefont{J.}~\bibnamefont{{Sapirstein}}},
  \bibinfo{journal}{Phys. Rev. A} \textbf{\bibinfo{volume}{37}},
  \bibinfo{pages}{307} (\bibinfo{year}{1988}).

\bibitem[{\citenamefont{Safronova et~al.}(1999)\citenamefont{Safronova,
  Johnson, and Derevianko}}]{relsd}
\bibinfo{author}{\bibfnamefont{M.~S.} \bibnamefont{Safronova}},
  \bibinfo{author}{\bibfnamefont{W.~R.} \bibnamefont{Johnson}},
  \bibnamefont{and}
  \bibinfo{author}{\bibfnamefont{A.}~\bibnamefont{Derevianko}},
  \bibinfo{journal}{Phys.\ Rev.\ A} \textbf{\bibinfo{volume}{60}},
  \bibinfo{pages}{4476} (\bibinfo{year}{1999}).

\bibitem[{\citenamefont{Safronova and Johnson}(2007)}]{review07}
\bibinfo{author}{\bibfnamefont{M.~S.} \bibnamefont{Safronova}}
  \bibnamefont{and} \bibinfo{author}{\bibfnamefont{W.~R.}
  \bibnamefont{Johnson}}, \bibinfo{journal}{Adv. At. Mol. Opt. Phys.}
  \textbf{\bibinfo{volume}{55}}, \bibinfo{pages}{191} (\bibinfo{year}{2007}).

\bibitem[{NIS({\natexlab{a}})}]{NIST1}
\bibinfo{note}{Yu. Ralchenko and A. E. Kramida and J. Reader and NIST ASD Team,
  \textit{NIST Atomic Spectra Database} (version 4.0). [Online]. Available:
  http://physics.nist.gov/asd. National Institute of Standards and Technology,
  Gaithersburg, MD}.

\bibitem[{\citenamefont{Moore}(1971)}]{NIST}
\bibinfo{author}{\bibfnamefont{C.~E.} \bibnamefont{Moore}},
  \emph{\bibinfo{title}{Atomic Energy Levels}}, vol.~\bibinfo{volume}{35} of
  \emph{\bibinfo{series}{Natl.\ Bur.\ Stand.\ Ref.\ Data Ser.}}
  (\bibinfo{publisher}{U.S.\ Govt.\ Print.\ Off.}, \bibinfo{year}{1971}).

\bibitem[{NIS({\natexlab{b}})}]{NIST2}
\bibinfo{note}{{J.} E. Sansonetti and W. C. Martin and S. L. Young, {\it
  Handbook of Basic Atomic Spectroscopic Data} (version 1.1.2). [Online]
  Available: http://physics.nist.gov/Handbook. (2005) National Institute of
  Standards and Technology, Gaithersburg, MD}.

\bibitem[{\citenamefont{Blundell et~al.}(1991)\citenamefont{Blundell, Johnson,
  and Sapirstein}}]{CC2}
\bibinfo{author}{\bibfnamefont{S.~A.} \bibnamefont{Blundell}},
  \bibinfo{author}{\bibfnamefont{W.~R.} \bibnamefont{Johnson}},
  \bibnamefont{and}
  \bibinfo{author}{\bibfnamefont{J.}~\bibnamefont{Sapirstein}},
  \bibinfo{journal}{Phys. Rev. A} \textbf{\bibinfo{volume}{43}},
  \bibinfo{pages}{3407} (\bibinfo{year}{1991}).

\bibitem[{\citenamefont{Safronova and Safronova}(2011)}]{ca}
\bibinfo{author}{\bibfnamefont{M.~S.} \bibnamefont{Safronova}}
  \bibnamefont{and} \bibinfo{author}{\bibfnamefont{U.~I.}
  \bibnamefont{Safronova}}, \bibinfo{journal}{Phys. Rev. A}
  \textbf{\bibinfo{volume}{83}}, \bibinfo{pages}{012503}
  (\bibinfo{year}{2011}).

\bibitem[{\citenamefont{Safronova et~al.}(2004)\citenamefont{Safronova,
  Williams, and Clark}}]{relsdrb}
\bibinfo{author}{\bibfnamefont{M.~S.} \bibnamefont{Safronova}},
  \bibinfo{author}{\bibfnamefont{C.~J.} \bibnamefont{Williams}},
  \bibnamefont{and} \bibinfo{author}{\bibfnamefont{C.~W.} \bibnamefont{Clark}},
  \bibinfo{journal}{Phys. Rev. A} \textbf{\bibinfo{volume}{69}},
  \bibinfo{pages}{022509} (\bibinfo{year}{2004}).

\bibitem[{\citenamefont{Sansonetti et~al.}(2011)\citenamefont{Sansonetti,
  Simien, Gillaspy, Tan, Brewer, Brown, Wu, and Porto}}]{Li}
\bibinfo{author}{\bibfnamefont{C.~J.} \bibnamefont{Sansonetti}},
  \bibinfo{author}{\bibfnamefont{C.~E.} \bibnamefont{Simien}},
  \bibinfo{author}{\bibfnamefont{J.~D.} \bibnamefont{Gillaspy}},
  \bibinfo{author}{\bibfnamefont{J.~N.} \bibnamefont{Tan}},
  \bibinfo{author}{\bibfnamefont{S.~M.} \bibnamefont{Brewer}},
  \bibinfo{author}{\bibfnamefont{R.~C.} \bibnamefont{Brown}},
  \bibinfo{author}{\bibfnamefont{S.}~\bibnamefont{Wu}}, \bibnamefont{and}
  \bibinfo{author}{\bibfnamefont{J.~V.} \bibnamefont{Porto}},
  \bibinfo{journal}{Phys. Rev. Lett.} \textbf{\bibinfo{volume}{107}},
  \bibinfo{pages}{023001} (\bibinfo{year}{2011}).

\bibitem[{\citenamefont{{Falke} et~al.}(2006)\citenamefont{{Falke}, {Tiemann},
  {Lisdat}, {Schnatz}, and {Grosche}}}]{falke}
\bibinfo{author}{\bibfnamefont{S.}~\bibnamefont{{Falke}}},
  \bibinfo{author}{\bibfnamefont{E.}~\bibnamefont{{Tiemann}}},
  \bibinfo{author}{\bibfnamefont{C.}~\bibnamefont{{Lisdat}}},
  \bibinfo{author}{\bibfnamefont{H.}~\bibnamefont{{Schnatz}}},
  \bibnamefont{and}
  \bibinfo{author}{\bibfnamefont{G.}~\bibnamefont{{Grosche}}},
  \bibinfo{journal}{Phys. Rev. A} \textbf{\bibinfo{volume}{74}},
  \bibinfo{pages}{032503} (\bibinfo{year}{2006}).

\bibitem[{\citenamefont{Steck}({\natexlab{a}})}]{dline85}
\bibinfo{author}{\bibfnamefont{D.~A.} \bibnamefont{Steck}},
  \bibinfo{note}{rubidium 85 D Line Data,” available online at
  http://steck.us/alkalidata (revision 2.1.4, 23 December 2010)}.

\bibitem[{\citenamefont{Steck}({\natexlab{b}})}]{dline87}
\bibinfo{author}{\bibfnamefont{D.~A.} \bibnamefont{Steck}},
  \bibinfo{note}{rubidium 87 D Line Data,” available online at
  http://steck.us/alkalidata (revision 2.1.4, 23 December 2010)}.

\bibitem[{com()}]{com}
\bibinfo{note}{Francesco Minardi, private communication}.

\bibitem[{\citenamefont{{Stellmer} et~al.}(2009)\citenamefont{{Stellmer},
  {Tey}, {Huang}, {Grimm}, and {Schreck}}}]{CWC-1}
\bibinfo{author}{\bibfnamefont{S.}~\bibnamefont{{Stellmer}}},
  \bibinfo{author}{\bibfnamefont{M.~K.} \bibnamefont{{Tey}}},
  \bibinfo{author}{\bibfnamefont{B.}~\bibnamefont{{Huang}}},
  \bibinfo{author}{\bibfnamefont{R.}~\bibnamefont{{Grimm}}}, \bibnamefont{and}
  \bibinfo{author}{\bibfnamefont{F.}~\bibnamefont{{Schreck}}},
  \bibinfo{journal}{Phys. Rev. Lett.} \textbf{\bibinfo{volume}{103}},
  \bibinfo{pages}{200401} (\bibinfo{year}{2009}).

\bibitem[{\citenamefont{{Desalvo} et~al.}(2010)\citenamefont{{Desalvo}, {Yan},
  {Mickelson}, {Martinez de Escobar}, and {Killian}}}]{sr-bec}
\bibinfo{author}{\bibfnamefont{B.~J.} \bibnamefont{{Desalvo}}},
  \bibinfo{author}{\bibfnamefont{M.}~\bibnamefont{{Yan}}},
  \bibinfo{author}{\bibfnamefont{P.~G.} \bibnamefont{{Mickelson}}},
  \bibinfo{author}{\bibfnamefont{Y.~N.} \bibnamefont{{Martinez de Escobar}}},
  \bibnamefont{and} \bibinfo{author}{\bibfnamefont{T.~C.}
  \bibnamefont{{Killian}}}, \bibinfo{journal}{Phys. Rev. Lett.}
  \textbf{\bibinfo{volume}{105}}, \bibinfo{pages}{030402}
  (\bibinfo{year}{2010}).

\bibitem[{\citenamefont{{Saffman} and {M{\o}lmer}}(2008)}]{ho1}
\bibinfo{author}{\bibfnamefont{M.}~\bibnamefont{{Saffman}}} \bibnamefont{and}
  \bibinfo{author}{\bibfnamefont{K.}~\bibnamefont{{M{\o}lmer}}},
  \bibinfo{journal}{Phys. Rev. A} \textbf{\bibinfo{volume}{78}},
  \bibinfo{pages}{012336} (\bibinfo{year}{2008}).

\bibitem[{\citenamefont{{McClelland} and {Hanssen}}(2006)}]{er1}
\bibinfo{author}{\bibfnamefont{J.~J.} \bibnamefont{{McClelland}}}
  \bibnamefont{and} \bibinfo{author}{\bibfnamefont{J.~L.}
  \bibnamefont{{Hanssen}}}, \bibinfo{journal}{Phys. Rev. Lett.}
  \textbf{\bibinfo{volume}{96}}, \bibinfo{pages}{143005}
  (\bibinfo{year}{2006}).

\bibitem[{\citenamefont{{Berglund} et~al.}(2007)\citenamefont{{Berglund},
  {Lee}, and {McClelland}}}]{er2}
\bibinfo{author}{\bibfnamefont{A.~J.} \bibnamefont{{Berglund}}},
  \bibinfo{author}{\bibfnamefont{S.~A.} \bibnamefont{{Lee}}}, \bibnamefont{and}
  \bibinfo{author}{\bibfnamefont{J.~J.} \bibnamefont{{McClelland}}},
  \bibinfo{journal}{Phys. Rev. A} \textbf{\bibinfo{volume}{76}},
  \bibinfo{pages}{053418} (\bibinfo{year}{2007}).

\bibitem[{\citenamefont{Tassy et~al.}(2010)\citenamefont{Tassy, Nemitz, Baumer,
  H\"{o}hl, Bat\"{a}r, and G\"{o}rlitz}}]{CWC-10a}
\bibinfo{author}{\bibfnamefont{S.}~\bibnamefont{Tassy}},
  \bibinfo{author}{\bibfnamefont{N.}~\bibnamefont{Nemitz}},
  \bibinfo{author}{\bibfnamefont{F.}~\bibnamefont{Baumer}},
  \bibinfo{author}{\bibfnamefont{C.}~\bibnamefont{H\"{o}hl}},
  \bibinfo{author}{\bibfnamefont{A.}~\bibnamefont{Bat\"{a}r}},
  \bibnamefont{and}
  \bibinfo{author}{\bibfnamefont{A.}~\bibnamefont{G\"{o}rlitz}},
  \bibinfo{journal}{J. Phys. B: At. Mol. Opt. Phys.}
  \textbf{\bibinfo{volume}{43}}, \bibinfo{pages}{205309}
  (\bibinfo{year}{2010}).

\bibitem[{\citenamefont{{Taglieber} et~al.}(2008)\citenamefont{{Taglieber},
  {Voigt}, {Aoki}, {H{\"a}nsch}, and {Dieckmann}}}]{three}
\bibinfo{author}{\bibfnamefont{M.}~\bibnamefont{{Taglieber}}},
  \bibinfo{author}{\bibfnamefont{A.}~\bibnamefont{{Voigt}}},
  \bibinfo{author}{\bibfnamefont{T.}~\bibnamefont{{Aoki}}},
  \bibinfo{author}{\bibfnamefont{T.~W.} \bibnamefont{{H{\"a}nsch}}},
  \bibnamefont{and}
  \bibinfo{author}{\bibfnamefont{K.}~\bibnamefont{{Dieckmann}}},
  \bibinfo{journal}{Phys. Rev. Lett.} \textbf{\bibinfo{volume}{100}},
  \bibinfo{pages}{010401} (\bibinfo{year}{2008}).

\bibitem[{\citenamefont{Vasilyev et~al.}(2002)\citenamefont{Vasilyev, Savukov,
  Safronova, and Berry}}]{beta}
\bibinfo{author}{\bibfnamefont{A.~A.} \bibnamefont{Vasilyev}},
  \bibinfo{author}{\bibfnamefont{I.~M.} \bibnamefont{Savukov}},
  \bibinfo{author}{\bibfnamefont{M.~S.} \bibnamefont{Safronova}},
  \bibnamefont{and} \bibinfo{author}{\bibfnamefont{H.~G.} \bibnamefont{Berry}},
  \bibinfo{journal}{Phys. Rev. A} \textbf{\bibinfo{volume}{66}},
  \bibinfo{pages}{020101} (\bibinfo{year}{2002}).

\end{thebibliography}

\end{document}